\documentclass[fleqn,twoside]{article}
\usepackage{espcrc2}

\usepackage{graphicx}
\usepackage[figuresright]{rotating}

\newcommand{\AmS}{{\protect\the\textfont2
  A\kern-.1667em\lower.5ex\hbox{M}\kern-.125emS}}

\title{Appearance of quark-hadron duality in the Rein-Sehgal model}

\author{Krzysztof M. Graczyk, Cezary Juszczak, Jan T. Sobczyk
\address[IFT]{Institute of Theoretical Physics \\
        University of Wroc\l aw\\
        pl. M. Borna 9, 50-204, Wroc\l aw, Poland}%
        \thanks{The authors were
supported by KBN grant 105/E-344/SPB/ICARUS/P-03/DZ211/2003-2005}
}
\begin{document}

\begin{abstract}
Quark-hadron duality in neutrino-nucleon reactions is investigated
under the assumption that cross sections in the resonance region
are given by the Rein-Sehgal model. The quantitative analysis of
the duality is done by means of appropriate integrals of the
structure functions in the Nachtmann variable. We conclude that
with the definition of the resonance region $W\in (M+m_{\pi}, 2$
GeV) the duality holds for neutrino-proton reaction $F_2$
structure function for $Q^2\in (0.5, 3)$ GeV$^2$ and it is absent
for neutrino-neutron reaction.
\end{abstract}

\maketitle

\section{INTRODUCTION}

The discovery of quark-hadron duality in electron-nucleon
interactions \cite{Duality} rises a natural question if the same
phenomenon can be seen also in neutrino-nucleon scattering.
Unfortunately the available data is not yet precise enough to
discuss the problem on the experimental level. At present the only
possibility is to analyze the existing theoretical models of
resonance production. This approach was adopted by Sato and Lee
\cite{LeeSato} who discovered that for their model of $\Delta$
production the quark-hadron duality is seen: the resonance peaks
of structure functions calculated at $Q^2=0.4,\ 1,\ 2,\ 4$ GeV$^2$
slide along the DIS structure functions (with CTEQ6 PDFs)
calculated at $10$ GeV$^2$, both as functions of the Nachtmann
variable.

In this contribution we wish to report an investigation of the
Rein-Sehgal (RS) model \cite{RS} of resonance production. The RS
model is used by almost all Monte Carlo generators of neutrino
interactions and understanding its properties is of practical
value. The model  describes resonance production in the region of
hadronic invariant mass $W$ below $2$ GeV by summing contributions
from 18 resonances. It contains also a non-resonant background
fine tuned in order to get a good agreement with the existing
single pion production (SPP) data.

We calculate the structure functions $F_1, F_2, F_3$ as they are
defined  by the RS model \cite{GJS}. They turn out to be linear
combinations of cross sections for three polarization states of
the intermediate $W$ boson. The aim of the RS model is to describe
only SPP channels and the elasticity coefficients are used in
order to select exclusive SPP channels from the overall resonance
production cross section. The non-resonant background is adjusted
only for SPP channels. The analysis of duality requires the
knowledge of the inclusive cross section in the resonance region.
In the spirit of the RS model one should add contributions from
inelastic channels taking care of complicated interference
patterns. In order to make the analysis simpler we introduce
1$\pi$ functions which are defined as probabilities that in the
given region of the kinematically allowed space the final state is
that of SPP. We calculate them numerically for each SPP channel
separately using the LUND fragmentation and hadronization
routines. All the details of our approach can be found in
\cite{GJS}.

\section{FORMALISM AND RESULTS}

\begin{figure}
\centerline{
\includegraphics[width=8cm]{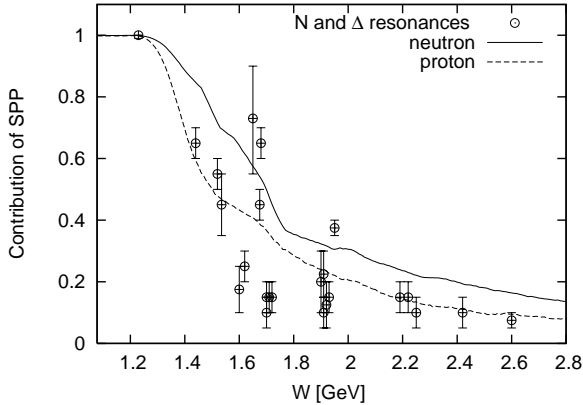}
}\caption{\small\small A comparison of 1$\pi$ functions defined in
Eq. \ref{1pion_def} with resonance elasticity factors taken from
PDG \cite{PDG}. In the case of $\nu$-neutron scattering we show
the sum of two functions corresponding to $\nu_{\mu} n\rightarrow
\mu^-p\pi^0$ and $\nu_{\mu} n\rightarrow \mu^-n\pi^+$ reactions
\label{elasticity}}
\end{figure}

The 1$\pi$ functions  are defined (for each exclusive channel
separately) as follows :
\begin{equation}
\label{1pion_def} f_{1\pi}(W,Q^2)\equiv \frac{\displaystyle
\frac{d^2 \sigma^{SPP}}{dWdQ^2}}{\displaystyle
\frac{d^2\sigma^{DIS}}{dWdQ^2}}.
\end{equation}

It turns out that  $f_{1\pi}(W,Q^2)$ do not depend on $Q^2$. Their
plots for CC $\nu N$ channels are presented in Fig.
\ref{elasticity}. In the case of neutron we show the sum of 1$\pi$
functions corresponding to two exclusive SPP channels. On the same
plot we show also the available data on elasticity of $N$ and
$\Delta$ resonances. We see that 1$\pi$ functions provide
satisfactory average description.

\begin{figure}
\centerline{
\includegraphics[width=8cm]{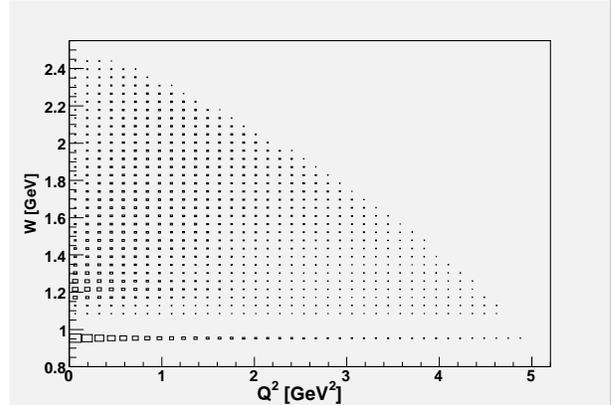}
}\caption{\small\small Double differential cross section
$\frac{d^2\sigma}{dWdQ^2}$ for neutrino energy $3$ GeV.
\label{slupki}}
\end{figure}

We  re-scale RS model structure functions  and obtain
approximations of structure functions in the RES region:
\begin{equation}
F_{i}^{RES}(x,Q^2) =
\frac{F^{RS}_{i}(x,Q^2)}{f_{1\pi}(x,Q^2)},\quad i=1,2,3.
\end{equation}

In our quantitative analysis we define the following function:
\begin{eqnarray}
\label{ratio_Q2=const} \mathcal{R}_2(Q^2;
Q^2_{DIS})=\frac{\displaystyle \displaystyle
\int_{\xi_{min}}^{\xi_{max}} d x F_2(\xi, Q^2)
}{\displaystyle\int_{\xi_{min}}^{\xi_{max}} d \xi F_2^{DIS}(\xi,
Q^2_{DIS})}.
\end{eqnarray}

$\xi$ is the Nachtmann variable.  $\xi_{max, min}$ are defined by
the conditions $W(\xi_{max}, Q^2)=M+m_{\pi}$ $W(\xi_{min},
Q^2)=W_{max}$. We investigate three choices for $W_{max}$. The
perfect duality would mean that $\mathcal{R}_2(Q^2;
Q^2_{DIS})\simeq 1$.

We select the physically relevant region in the $Q^2$ variable as
$Q^2\leq 3$~GeV$^2$ by looking at double differential cross
section $\frac{d^2\sigma}{dWdQ^2}$ for neutrino energy in a few
GeV region, see Fig. 2 \cite{JNS}. Having in mind MC generators of
events the quark-hadron duality should ensure the smooth passage
of the cross section for RES and DIS contributions in the relevant
kinematical transition region.

\begin{figure}
\centerline{
\includegraphics[width=8cm]{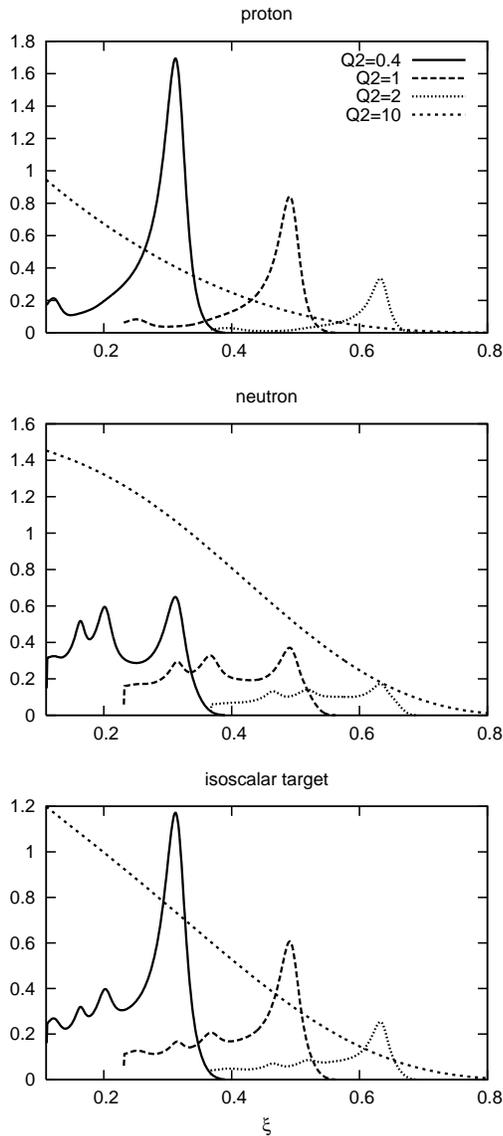}
}\caption{\small\small Manifestation of quark-hadron duality for
$F_2$ structure function for CC neutrino reactions on proton,
neutron and isoscalar target. $F_2$ for RS model is calculated at
$Q^2=0.4,\ 1,\ 2$ GeV$^2$ and for DIS at $10$ GeV$^2$.
\label{slizganie}}
\end{figure}

In Fig.~\ref{slizganie} we show  $F_2$ structure functions for CC
reactions defined by the RS model for three values of $Q^2$:
$0.4,\  1,\  2$ GeV$^2$ and also DIS $F_2$ structure functions
calculated at $Q^2=10$ GeV$^2$. The targets are from the top to
the bottom: proton, neutron and their average isoscalar one. The
plots for the DIS part are done with GRV94 PDFs \cite{GRV94}. RS
structure functions are yet not re-scaled by 1$\pi$ functions.
Similar plots for $F_2$ at $Q^2=0.5,~1$ GeV$^2$ were shown before
in \cite{Lipari}. We see that in the case of proton and isoscalar
target typical manifestation of local quark-hadron duality: the
$\Delta$ peak slides along the DIS curve. The duality does not
seem to apply neither to other resonances nor to neutron target
even in the case of the prominent $\Delta$ resonance.

\begin{figure}
\centerline{
\includegraphics[width=6.5cm]{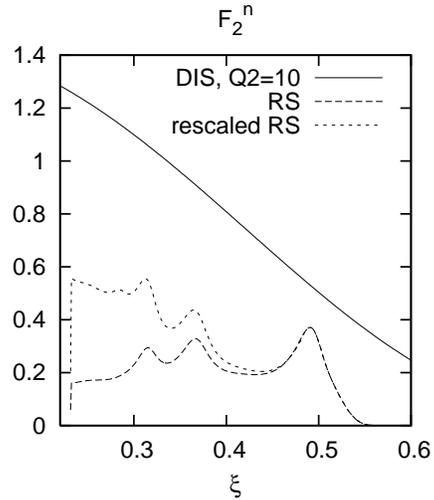}
}\caption{\small \small The comparison of $F_2$ structure
functions for the RS model, re-scaled RS model and the DIS. The
structure functions are calculated for CC $\nu$-neutron scattering
at $Q^2$=1 GeV$^2$. \label{1pion_in_action}}
\end{figure}

In Fig.~\ref{1pion_in_action} we show the role played by 1$\pi$
functions. On the same plot both Rein-Sehgal and 1$\pi$ function
re-scaled RES $F_2$ structure functions for CC reaction on neutron
are presented. We see that the modifications apply mostly to the
region of $W$ close to 2~GeV.

\begin{figure}
\centerline{
\includegraphics[width=8cm]{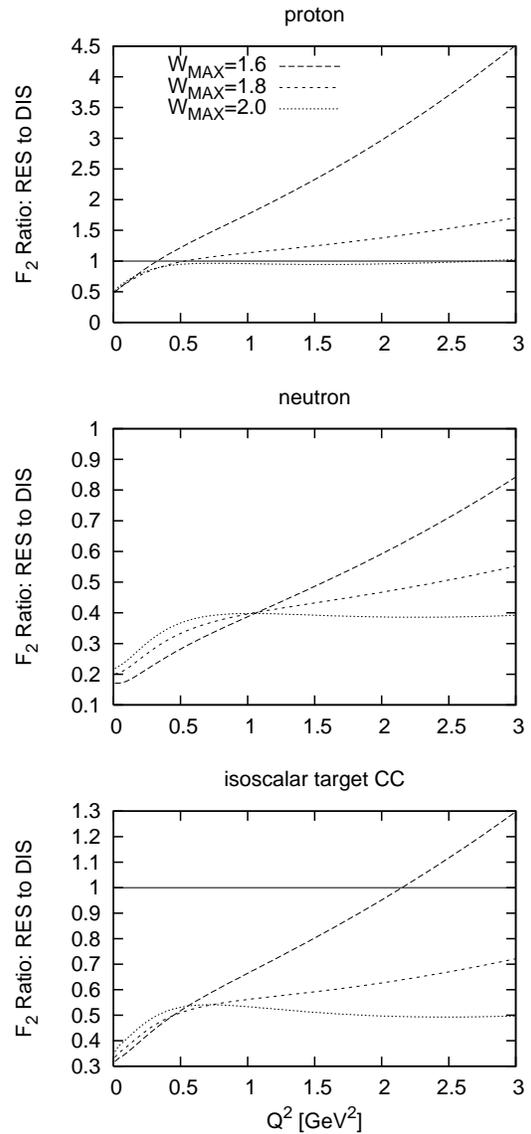}
}\caption{\small\small Function $\mathcal{R}_2(Q^2; Q^2_{DIS}=10$
GeV$^2$) (see Eq.~\ref{ratio_Q2=const}) for CC reaction on proton,
neutron and isoscalar targets with $W_{max}= 1.6,\ 1.8,\ 2.0$ GeV.
\label{ratios}}
\end{figure}

Fig.~\ref{ratios} presents the functions defined in
Eq.~\ref{ratio_Q2=const}  for three different integration regions
i.e. for three definitions of the resonance region: $W\in
(M+m_{\pi}, W_{max})$ where $W_{max}=1.6,\  1.8,\  2$ GeV. We see
that the choice $W_{max}=2$ GeV makes the functions slowly varying
in the wide region of $Q^2$. The behavior for small values of
$Q^2\leq 0.5$~GeV$^2$ is very different in agreement with the
predictions made in \cite{CI}. For the proton target the best
choice is $W_{max}=2$~GeV while for other two targets it is
preferable to choice the resonance region as more confined at the
price of significant variations with $Q^2$. The choice
$W_{max}=2$~GeV is a natural one for the RS model. But it is
suggested in \cite{Bodek} that the RS model underestimates the
cross section for $W\geq 1.7$~GeV. We conclude that with the
choice $W_{max}=2$~GeV the duality holds very well for the proton
and badly for other targets. It seems to be difficult to have
duality for all the targets simultaneously.

\vskip 0.3truecm\noindent
{\bf Acknowledgments}\\

We thank Olga Lalakulich for a discussion on resonance
elasticities and Jaros\l aw Nowak for preparing for us the Figs.
\ref{elasticity} and \ref{slupki}.

\end{document}